\documentclass[useAMS,usenatbib,fleqn]{mnras}
\usepackage{amsopn, natbib} \usepackage[utf8]{inputenc}
\usepackage{amsmath} \usepackage{caption} \usepackage{amsfonts}
\usepackage{amssymb} \usepackage{graphicx} \usepackage{hyperref}
 \usepackage{psfrag}
\usepackage{tabularx}
\usepackage[dvipsnames]{xcolor}
\newcommand{\U}{\mathbfit{U}}

\newcommand{\ppk}{\mathbfit{k}_{\perp}} 
\newcommand{\kpar}{k_{\parallel}}
\newcommand{\HI}{{H{\sc I}~}}
\title[Cosmic Dawn 21-cm Power Spectrum using uGMRT]{On the Prospects of 
Measuring the Cosmic Dawn 21-cm Power Spectrum using the Upgraded Giant 
Meterwave Radio Telescope (uGMRT)}

\author[Chatterjee et al.]{Suman Chatterjee$^{1,2}$\thanks{E-mail:
    suman05@phy.iitkgp.ernet.in}, Somnath
  Bharadwaj$^{1,2}$\thanks{E-mail: somnath@phy.iitkgp.ernet.in }
  \\ $^{1}$Department of Physics, Indian Institute of Technology
  Kharagpur, Kharagpur - 721 302, India.\\ $^{2}$Centre for
  Theoretical Studies, Indian Institute of Technology Kharagpur,
  Kharagpur - 721 302, India.}
\begin{document}
\date{\today}

\pagerange{\pageref{firstpage}--\pageref{lastpage}} \pubyear{2018}

\maketitle
\label{firstpage}
\begin{abstract}
A recent observation by the EDGES collaboration 
shows a strong absorption signal in the global 21-cm spectrum from
around a redshift of z = 17. This absorption is stronger than the
maximum prediction by existing models and indicates that the spatial
fluctuations of the \HI 21-cm brightness temperature at Cosmic Dawn
could be an order of magnitude larger than previously expected. 
In this  paper, we investigate the
prospects of detecting the \HI 21-cm power spectrum from Cosmic Dawn
using uGMRT. We find that a $10\sigma$ detection of the enhanced \HI
21-cm signal power spectrum is possible within $70, 140$ and $400$
Hours of observation for an optimistic, moderate and pessimistic
scenario respectively, using the Band-1 of uGMRT.  
This could  be an
useful probe of the interaction between the baryon and dark matter
particles in the early universe.
 We also present a 
comparison of  the uGMRT predictions with those  for the  future SKA-Low.
\end{abstract}

\newpage

\section{Introduction} Observations with the Experiment to Detect the
Global Epoch of Reionization Signature (EDGES) have recently resulted
in the detection of an absorption profile with full width half maxima
(FWHM) $19 \, {\rm MHz}$ centred at $78 \, {\rm MHz}$ in the sky
averaged spectrum of the background radiation in the frequency range
$50-100 \, {\rm MHz}$ \citep{Bowman2018}. If confirmed by other
similar experiments like the Large-Aperture Experiment to Detect the
Dark Ages (LEDA; \citealt{Bernardi2016}) , the Sonda Cosmol\'ogica de
las Islas para la Detecci\'on de Hidr\'ogeno Neutro (SCI-HI;
\citealt{Voytek2014}), the Probing Radio Intensity at high z from
Marion (PRIZM; \citealt{Philip2018}) and the Shaped Antenna measurement of the background
Radio Spectrum 2 (SARAS 2; \citealt{Singh2017}), this can be
interpreted as the neutral Hydrogen (\HI) 21-cm absorption profile
resulting from the Lyman-$\alpha$ coupling due to the formation of the
first stars in the early universe \citep{Pritchard2012}.  However, the
observation indicates a dip with amplitude $0.5 \, {\rm K}$ which is
more than a factor of two larger than the largest predictions
\citep{Cohen2017}. \citet{Barkana2018} have proposed that it is
possible to explain this enhanced dip through a possible interaction
between the baryons and dark matter particles (b-DM interaction). Such
models \citep{Barkana2018, Fialkov2018} have predicted a $10$ fold
enhancement of the spatial fluctuations of the redshifted \HI 21-cm
brightness temperature $\delta T_b(x)$. We note that other alternative
explanations have also been proposed \citep{Ewall-Wice2018, Feng2018}
to explain the enhanced dip. The latter models incorporate an
enhancement in the radio background and they do not predict such
enhancement in $\delta T_b(x)$.

\citet{Dvorkin2014},\citet{Tashiro2014},\citet{Munoz2015} and \citet{Xu2018}
  have considered b-DM interaction in the context of cosmology and
  large-scale structure formation. 
   \citet{Barkana2018}  have  assumed a non-standard
  Coulomb-like interaction between dark-matter particles and baryons
  that does not depend on whether the baryons are free or bound within
  atoms. By combining this with the radiation
  emitted by the first stars during cosmic dawn,  they find a  strong 21-cm absorption that can
  explain the feature measured by EDGES. These  models \citep{Barkana2018, Fialkov2018} also 
  predict a 10 fold enhancement of the spatial fluctuations of the
  redshifted HI 21-cm brightness temperature $\delta T_b(x)$.
  \citet{Munoz2018a} explore whether dark-matter
  particles with an electric ``minicharge" can significantly alter the
  baryonic temperature and affect the global 21-cm signal. They
  find that the entire  dark matter cannot be minicharged at
  a significant level, the constraints coming from  Galactic and
  extragalactic magnetic fields. However,  minicharged particles that 
 comprise a subpercent fraction of the dark matter and have a charge
  $\sim 10^{-6}$ in units of the electron charge and masses $m_{\chi}
  \sim 1 - 60 \, {\rm MeV}$ can significantly cool down the
  baryons and explain the  EDGES result while remaining consistent
  with other observational constraints. In a recent paper,
  subsequent to the submission of the present {\em paper}, 
  \citet{Munoz2018b} have analysied the 21-cm brightness temperature 
  fluctuations for the minicharge model. Their study confirms a significant
  enhancement in the predicted 21-cm brightness temperature 
  fluctuations, including an increase in the amplitude
  of the baryon acoustic oscillation.  

Upcoming experiments such as the Hydrogen Epoch of Reionization Array
(HERA; \citealt{DeBoer2017}) and the Square Kilometre Array (SKA;
\citealt{Koopmans2015}) have the potential of measuring the \HI 21-cm
power spectrum from Cosmic Dawn $(50 - 100 \, {\rm MHz})$. Both of
these experiments should easily be able to measure the corresponding
enhanced \HI 21-cm power spectrum predicted by the b-DM interaction
models \citep{Barkana2018}.

The Giant Meterwave Radio Telescope (GMRT; \citealt{Swarup1991}) is
one of the largest and most sensitive fully operational low-frequency
radio telescopes in the world today. The array configuration of $30$
antennas (each of $45 \, {\rm m}$ diameter) spanning over $25 \, {\rm
  km}$, provides a total collecting area of about $30,000 \, {\rm
  sq. \, m}$ at metre wavelengths. The GMRT is being upgraded (uGMRT,
\citealt{Gupta2017}) to have seamless frequency coverage, as far as
possible, from $50$ to $1500 \, {\rm MHz}$. Band-1 of uGMRT, which is
yet to be implemented, is expected to cover the frequency range $50 -
80 \, {\rm MHz}$.  Earlier \citet{Shankar2009} envisaged a $50 \, {\rm
  MHz}$ system developed for GMRT to provide imaging capability in the
frequency range $30-90 \, {\rm MHz}$.

In this paper, we investigate the prospects of detecting the
redshifted \HI 21-cm signal power spectrum from Cosmic Dawn using the
uGMRT. For the purpose of this analysis we have considered a
functional bandwidth of $B = 20 \, {\rm MHz}$ centred at $\nu_c = 78
\, {\rm MHz}$, consistent with the frequency coverage described in
\citet{Shankar2009}. Frequencies above $90 \, {\rm MHz}$ are used for
FM transmission which restricts the allowed frequency range.
We have also carried out a similar 
analysis for the future SKA-Low, and we present a comparison of 
the predictions  for the  uGMRT with those for the future 
SKA-Low.

\section{Methodology} We have simulated the uGMRT baseline configuration
for $8 \, {\rm hours}$ of observation targeted on a field at
$+60^{\circ}$ DEC with $16 \, {\rm s}$ integration time.
The entire analysis has been restricted to 
  baselines with antenna separations within $2\, {\rm km}$,
 which contains the bulk of the cosmological signal. We assume
that the bandwidth $B = 20 \, {\rm MHz}$ is divided into $N_c = 200$
spectral channels of $\Delta \nu_c = 100 \, {\rm KHz}$. Note that the
values of $B$, $N_c$ and $\Delta \nu_c$ assumed here are only
representative values, and the actual values in the final
implementation of the telescope may be somewhat different.
The $20 \, {\rm MHz}$   bandwidth 
spans the redshift range  $z=15$ to $20$, and the \HI 21-cm power spectrum 
may evolve significantly within this redshift range. Consequently  
we have considered three  bands each of 
width $6 \, {\rm MHz}$ centred at 
$72 \, {\rm MHz} $, $78 \, {\rm MHz}$ and $84 \, {\rm MHz}$ which
correspond to $z = 18.7, 17.2$ and $15.9$ respectively. We assume that
the measurements from these three bands are combined to enhance the
signal-to-noise ratio (SNR).

Considering  the simulated baseline $u-v$ distribution, we use
$\mathbfit{k}_{\perp}=2\pi\mathbfit{U}/r$ and $k_{\parallel m}=2\pi
m/r^{'} B$ to estimate the Fourier modes at which the brightness
temperature fluctuations $\Delta T(\mathbfit{k})$ will be measured by
this observation. Here $\mathbfit{U}$ refer to different baselines, $0
\leq m \leq N_c/2$, $r$ is the co-moving distance corresponding to
$\nu_c$ and $r^{'} = dr/d\nu$ evaluated at $\nu = \nu_c$. 
We assume that the measured $\Delta T(\mathbfit{k})$ values
are gridded in $(\mathbfit{k}_{\perp},k_{\parallel})$ space 
with a grid spacing  $\Delta k_{\perp}=2\pi D/ \lambda_c r$ and 
$\Delta k_{\parallel}=2\pi /r^{'} B$, and the gridded values 
are used to estimate  the power spectrum $P(\mathbfit{k}_g)$
at each grid point $\mathbfit{k}_g$. Here we have used the 
simulation to estimate 
the sampling function $\tau(\mathbfit{k}_g)$ which refers to the number of 
 distinct $\Delta T(\mathbfit{k})$
measurements that contribute to each grid point $\mathbfit{k}_g$.

For the Cosmic Dawn \HI 21-cm signal we have used the
  value of the dimensionless \HI 21-cm power spectrum $\Delta^2_{\rm
    \HI}(k) = k^3 P_{{\rm \HI}}(k)/2 \pi^2$ from
  earler works \citep{Santos2010,Mellema2013},  where $P_{{\rm \HI}}(k)$ refers 
  to the \HI 21-cm power spectrum.
\citep{Barkana2018,Fialkov2018}. We have used the 
  $z=17$ \HI 21-cm power spectrum  
predictions from \citet{Santos2010} as the fiducial model for
  all the three bands which we have considered here.
These values correspond to the standard scenario, we expect a $10$
fold enhancement in the brightness temperature fluctuations {\it i.e.}
a dimensionless power spectrum of $100\Delta^2_{\rm \HI}$ in the
presence of the b-DM interaction

In addition to the \HI 21-cm power spectrum, 
we also have the noise power spectrum $P_N(\ppk, \kpar)$ which can be 
estimated as follows. The measured  power spectrum  is related to the observed 
visibilities $V(\U,\nu)$ as 
\begin{equation}
P(\ppk, \kpar) = \frac{r^{'}}{\tilde{Q}} \int \langle 
V(\U, \nu) V^{*}(\U, \nu + \Delta \nu) \rangle  d(\Delta\nu)
\label{eq:pk}
\end{equation}
which can be obtained from (eq.~15.) of \citet{Bharadwaj2005},
where $\tilde{Q} = (\partial B/\partial T)^2 r^{-2} \int A(\theta)^2 d^2\theta$, $A(\theta)$ is the primary beam pattern of the telescope, $(\partial B/\partial T) = 2k_B/\lambda^2$,
Considering the noise contribution $N(\U,\nu)$
to the observed visibilities $V(\U,\nu)$, and assuming that the noise at two 
different frequency channels is uncorrelated, 
we have the noise power spectrum contribution from the visibilities 
measured at a single baseline $\U$ to be 
\begin{equation}
P_N(\ppk, \kpar)=\frac{r^{'} \langle \mid N(\U,\nu)  \mid^2 \rangle \, 
\Delta \nu_c}{\tilde{Q}}
\label{eq:x1}
\end{equation}
which is independent of $\kpar$. The real and imaginary 
parts of $N(\U,\nu)$ both have equal variance $\sigma_N^2$  
\citep{Chengalur2007} with 
\begin{equation}
\sigma_N^2 = \frac{2}{N_p \Delta \nu_c \, \Delta t}
\left(\frac{k_B T_{sys}}{ \eta A_g} \right)^2 
\label{eq:x2}
\end{equation}
 where $T_{sys}$ is the system temperature, 
$\Delta t$ is the integration time, $N_p$ is the number of 
polarizations and the antenna efficiency $\eta$ is defined through 
 $\lambda^2/ \eta \, A_g = 
\int A(\theta) d^2 \theta $ where $A_g$ is the geometrical 
area of the antennas. 

In our analysis we have gridded  
the simulated baseline distribution  on to  the $(\ppk,\kpar)$ grid 
introduced earlier. Figure  5 of \citealt{Choudhuri2014} shows the 
baseline distribution corresponding  to the uGMRT observations
 considered here. Note that the $\U$ values need to be multiplied 
 by a factor of $\approx 2$ to scale them from $150 \, {\rm MHz}$
 \citep{Choudhuri2014} to the central frequency $78 \, {\rm MHz}$
 considered here.
 We see that the simulated baselines do not uniformly 
sample the $u-v$ (or equivalently $\ppk$) plane, and we have used the  
grid sampling function  $\tau(\mathbfit{k}_g)$ to quantify the number 
of baselines which contribute to each grid point $\mathbfit{k}_g$.
Incorporating 
this, the noise power spectrum at each grid point can be expressed as 
\begin{equation}
P_N(\mathbfit{k}_g) = \frac{T_{sys}^2 \, r^{'} \, r^2 }
{\Delta t \, N_p \, \tilde{\eta} \, \tau(\mathbfit{k}_g)} \, ,
\label{eq:noise}
\end{equation}
where we have defined the dimensionless factor 
  $\tilde{\eta} = [\int A^2(\theta) d^2\theta]/[\int A(\theta) d^2\theta]^2$.

We have assumed the system temperature $T_{sys} = 3000 \, {\rm K}$,
 number of polarizations $N_p = 2$ and integration time $\Delta t = 16 \, {\rm s}$.
At the frequencies of our interest the uGMRT primary beam pattern
is well approximated by a Gaussian
$A(\theta)=e^{-(\theta/\theta_0)^2}$ with $\theta_0=3.1^{\circ}$
whereby $\tilde{\eta}=54.4$. Note that it is well justified to use the 
flat sky approximation for the uGMRT. The analysis till now considers $8$ hours 
of observation which roughly corresponds to a single night. For longer observations, 
the noise power  spectrum has been scaled inversely with the number of observing 
nights.

We have binned the $k$-range accessible to uGMRT into $10$
 logarithmic bins. For each bin $a$,we the binned power spectrum 
$P_a=\sum_g w_g P(\mathbfit{k}_g)$ is a weighted sum  of  the 
power spectrum measured at all the  grid points within the bin. 
The weights have been chosen as $w_g= A P_{{\rm
    \HI}}(\mathbfit{k}_g)/[P_{{\rm \HI}}(\mathbfit{k}_g) +
  P_N(\mathbfit{k}_g)]^2$ (with normalization constant $A$) to
optimise the SNR. We use the variance $(\Delta P_a)^2 = {1/
  \sum_{g} [P_{{\rm \HI}}(\mathbfit{k}_g) + P_N(\mathbfit{k}_g)]^{-2}}$
to quantify the uncertainty with which it will be possible to measure  
power spectrum in each bin.

Foreground removal (e.g. \citealt{Ali2008}) is an important 
issue for detecting the cosmological  21-cm power spectrum. 
Several studies have  
shown that the foregrounds are expected to be confined to a wedge 
which is approximately bounded by 
\begin{equation}
k_{\parallel} \le \left[ \frac{r \, \sin(\theta_l)}{r_{'} \, \nu_c} \right] 
\,k_{\perp} 
\label{eq:a1}
\end{equation}
in the $(k_{\parallel},k_{\perp})$ plane \citep{Datta2010,Vedantham2012,Morales2012,
Parsons2012b, Trott2012} where   $\theta_l$  refers to the largest 
angle (relative to the telescope's pointing direction) 
from which  we have a  significant foreground contamination.  
Only  the $k$-modes outside this 
``foreground wedge" can be used for power spectrum estimation. 
The exact extent of this wedge is however still debatable 
(see \citealt{Pober2014} for a detailed discussion), and  
for the  purpose of this work we consider three different cases 
which differ in the extent of the foreground wedge.

\begin{itemize}
\item {\it \textbf{Case I}:} This is the most optimistic scenario,
  where we assume that the foregrounds have been removed perfectly and
  the whole $\bf{k}$ space accessible by uGMRT is available for
  measuring the \HI 21-cm signal.

\item {\it \textbf{Case II}:} In this moderate scenario we assume that
  the foreground contributions from angles beyond $\theta_l=18^{\circ}$ from
  the center of the field of view are highly suppressed by tapering
  the sky response \citep{Choudhuri2014}.  Note that the first null of
  the uGMRT primary beam pattern at $78 \, {\rm MHz}$ is expected at
  $\sim 6^{\circ} $. In this case the Fourier modes $\kpar \leq 1.813
  \, |\ppk|$ are foreground contaminated
(eq. \ref{eq:a1}), and only the modes outside 
  this foreground wedge are used for measuring the \HI 21-cm signal.

\item {\it \textbf{Case III}:} In this pessimistic scenario we assume
  that the foreground contribution extends till the horizon  
$(\theta_l=90^{\circ})$ , and the
  Fourier modes $\kpar \leq 5.964 \, |\ppk|$ are foreground
  contaminated (eq.\ref{eq:a1}). Only the modes outside this foreground wedge are used 
  for measuring the \HI 21-cm signal.

\end{itemize}

\section{Result}
We first consider very large observing times for which 
$P_N \rightarrow 0$, and the $1-\sigma$ error $\Delta P_a$ on the measurement
 of  $P_a$ converges to the cosmic variance. We find
  that ${\rm SNR} > 5$ can be achieved at $k > 0.02 {\rm Mpc}^{-1}$,
  $0.04 {\rm Mpc}^{-1}$ and $0.1 {\rm Mpc}^{-1}$ for Case I, Case II
  and Case III respectively. We only consider these $k$-modes for our
subsequent analysis. We see that in all the three cases there is a
reasonably large $k$-range where a detection is possible provided we
have sufficiently deep observations.

\begin{figure*}
\psfrag{optimistic}{Case \, I} \psfrag{Modarate}{Case \, II}
\psfrag{pessimistic}{Case \, III} \psfrag{100DHI2}{$100\Delta^2_{\rm
    \HI}$} \psfrag{10DHI2}{$10\Delta^2_{\rm \HI}$}
\psfrag{DHI2}{$\Delta^2_{\rm \HI}$} \psfrag{HI signal}{\qquad \HI
  signal} \psfrag{100 Hr}{100 Hours} \psfrag{500 Hr}{500 Hours}
\psfrag{1000 Hr}{1000 Hours} \psfrag{D2kk-mK2}{$\Delta^2(k)\, {\rm
    mK}^2$} \psfrag{k-Mpc-1}{$k \, \rm{Mpc}^{-1}$} \centering
\includegraphics[scale=0.75, angle = 270]{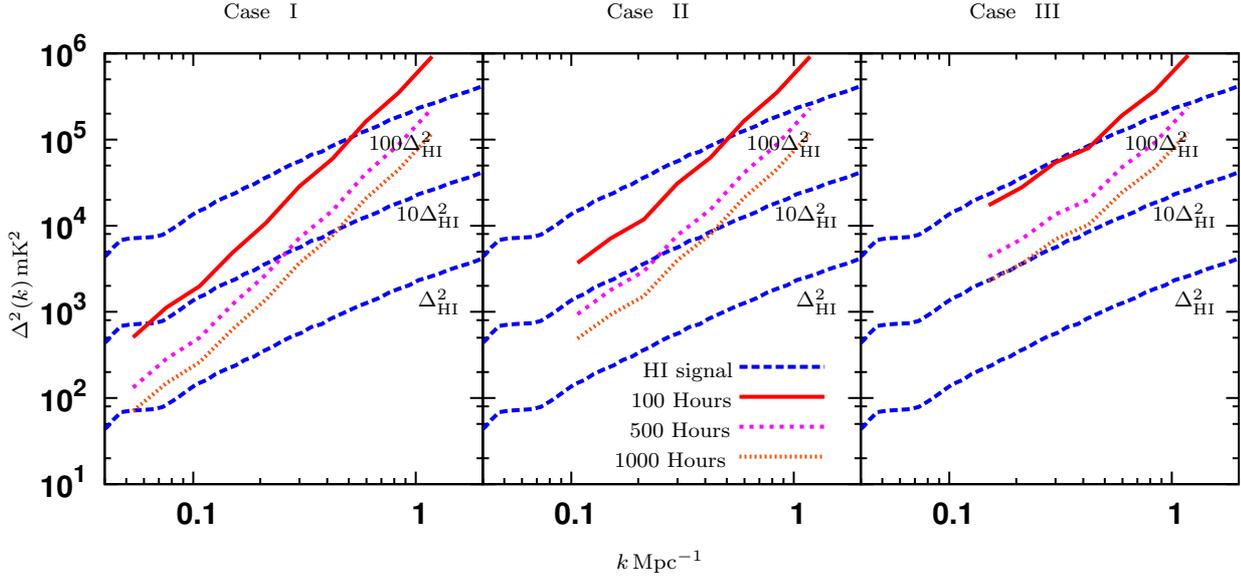}
\caption{This shows a comparison of the dimensionless 
\HI 21-cm signal power
  spectrum and the corresponding $1-\sigma$ error  for an uGMRT observation. The
  left, middle and right panels show results for Case I, II and III
  respectively. The \HI signal is shown in dashed lines (as mentioned
  in the figure). The error on the measurement of the power spectrum for $100$, $500$ and $1000$
  hours is shown in solid, dotted and fine-dotted lines respectively.}
\label{fig:b}
\end{figure*}

The system noise dominates $\Delta P_a$ for small observing 
times. Figure~\ref{fig:b} shows a comparison between the dimensionless 
\HI 21-cm signal power spectrum and  the corresponding  $1-\sigma$ error.   
 For Case I
we find that $100\Delta^2_{\rm \HI}$ and $10\Delta^2_{\rm \HI}$ can be
detected with $100$ and $500$ hours of observation for Fourier modes 
$0.06 < k < 0.5 \, {\rm Mpc}^{-1}$ and $0.06 < k< 0.25 \, {\rm
  Mpc}^{-1}$ respectively. However, a detection of $\Delta^2_{\rm
  \HI}$ will require more than $1000$ hours of observation. In Case
II, it is possible to detect $100\Delta^2_{\rm \HI}$ and
$10\Delta^2_{\rm \HI}$ in the $k$-range, $0.1 < k < 0.45 \, {\rm
  Mpc}^{-1}$ and $0.1 < k< 0.25 \, {\rm Mpc}^{-1}$ in $100$ and $500$
hours of observation respectively. For the pessimistic scenario, {\it
  i.e.} Case III, we find that $100\Delta^2_{\rm \HI}$ can be detected in $100$
 hours of observation in the $k$-range $0.1 < k < 0.4 \, {\rm Mpc}^{-1}$ and  
the detection of $10\Delta^2_{\rm \HI}$ will require more than $1000$ hours of
observation.

 We finally consider the situation where all the
  available $k$-modes are combined for a detection of the \HI 21-cm
  signal.  Here we have a single parameter $A_{{\rm \HI}}$ which is
  the amplitude of the \HI 21-cm power spectrum. We have estimated the
  ${\rm SNR}$ for the measurement of $A_{{\rm \HI}}$ using
\begin{equation}
{\rm SNR}^2 = \frac{1}{2} \sum_{\mathbfit{k}_g} \left[\frac{\partial
    P_{{\rm \HI}}(\mathbfit{k}_g)}{\partial ln A_{{\rm \HI}}}\right]^2
[P_{{\rm \HI}}(\mathbfit{k}_g)+ P_N(\mathbfit{k}_g)]^{-2} \, .
\label{eq:snr}
\end{equation}

\begin{figure*}
\psfrag{optimistic}{Case \, I} \psfrag{moderate}{Case \, II}
\psfrag{pessimistic}{Case \, III} \psfrag{100DHI2}{$100\Delta^2_{\rm
    \HI}$} \psfrag{10DHI2}{$10\Delta^2_{\rm \HI}$}
\psfrag{DHI2}{$\Delta^2_{\rm \HI}$} \psfrag{SNR}{SNR}
\psfrag{Observation Hours}{Observation Hours} \centering
\includegraphics[scale=0.75, angle = 270]{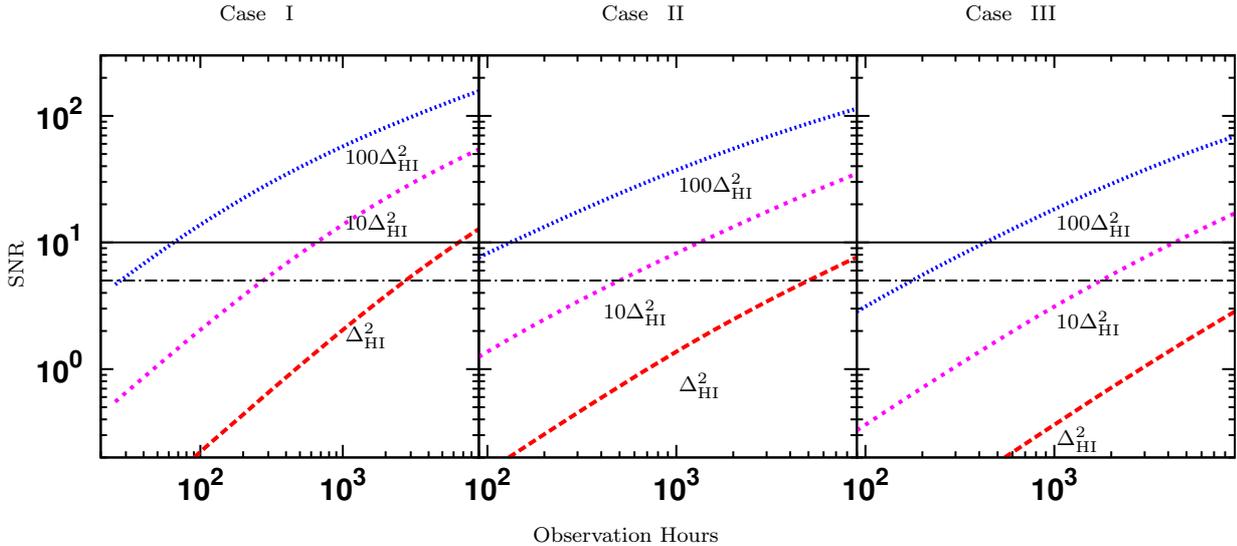}
\caption{The left, middle and right panels show the predictions for
  Case I, II and III respectively when all the available $k$-modes are
  combined. The dashed, dotted and fine-dotted lines show the
  predictions for the \HI 21-cm signal power spectrum $\Delta^2_{\rm
    \HI}$, $10\Delta^2_{\rm \HI}$ and $100\Delta^2_{\rm \HI}$
  respectively . The horizontal dot-dashed and solid lines mark the
  SNR of $5$ and $10$ respectively.}
\label{fig:c}
\end{figure*}

Figure~\ref{fig:c} shows the predicted SNR as a function of the observing time.
 The
horizontal solid and the dot-dashed lines mark the SNR value of $10$
and $5$ respectively.  For Case I we find that a $10\sigma$ detection
of $100\Delta^2_{\rm \HI}$, $10\Delta^2_{\rm \HI}$ and $\Delta^2_{\rm
  \HI}$ is possible in $\sim 70,\, 700 \, {\rm and} \, 6000$ hours of
observation respectively. A $5\sigma$ detection of $\Delta^2_{\rm
  \HI}$ can be achieved in $~3000$ hours of observation.  In Case II
and Case III, it takes $\sim 140, 1400$ hours and $\sim 400, 4000$
hours for a $10\sigma$ detection of $100\Delta^2_{\rm \HI}$ and
$10\Delta^2_{\rm \HI}$ respectively. It is not possible to detect
$\Delta^2_{\rm \HI}$ within reasonable observation time when we
consider the pessimistic scenario.

\begin{figure*}
\psfrag{optimistic}{Case \, I} \psfrag{Modarate}{Case \, II}
\psfrag{pessimistic}{Case \, III} \psfrag{100DHI2}{$100\Delta^2_{\rm
    \HI}$} \psfrag{10DHI2}{$10\Delta^2_{\rm \HI}$}
\psfrag{DHI2}{$\Delta^2_{\rm \HI}$} \psfrag{HI signal}{\qquad \HI
  signal} \psfrag{100 Hr}{100 Hours} \psfrag{500 Hr}{500 Hours}
\psfrag{1000 Hr}{1000 Hours} \psfrag{D2kk-mK2}{$\Delta^2(k)\, {\rm
    mK}^2$} \psfrag{k-Mpc-1}{$k \, \rm{Mpc}^{-1}$} \centering
\includegraphics[scale=0.75, angle = 270]{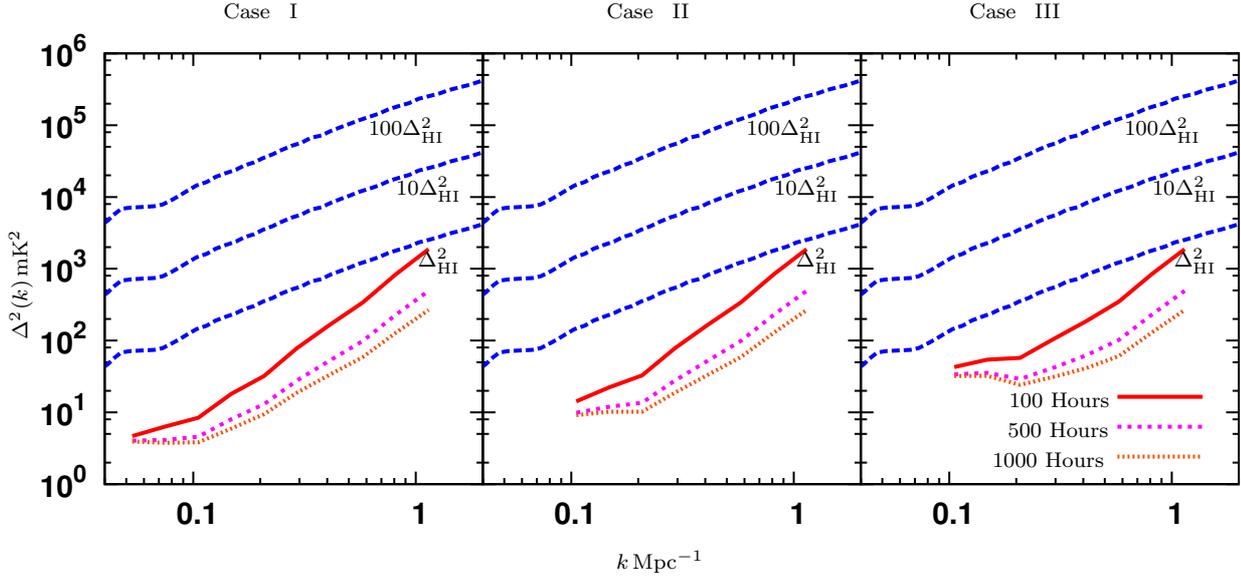}
\caption{This shows a comparison of the dimensionless 
\HI 21-cm signal power spectrum and the corresponding $1-\sigma$ error  for an SKA-Low observation. The
  left, middle and right panels show results for Case I, II and III
  respectively. The \HI signal is shown in dashed lines (as mentioned
  in the figure). The error on the measurement of the power spectrum for $100$, $500$ and $1000$
  hours is shown in solid, dotted and fine-dotted lines respectively.}
\label{fig:d}
\end{figure*}

\begin{figure}
\psfrag{opt}{Case \, I} 
\psfrag{mod}{Case \, II}
\psfrag{pess}{Case \, III} 
\psfrag{100DHI2}{$100\Delta^2_{\rm\HI}$} 
\psfrag{10DHI2}{$10\Delta^2_{\rm \HI}$}
\psfrag{DHI2}{$\Delta^2_{\rm \HI}$} 
\psfrag{SNR}{SNR}
\psfrag{Observation Hours}{Observation Hours} 
\psfrag{SKA-low}{ }
\centering
\includegraphics[scale=0.45, angle = 270]{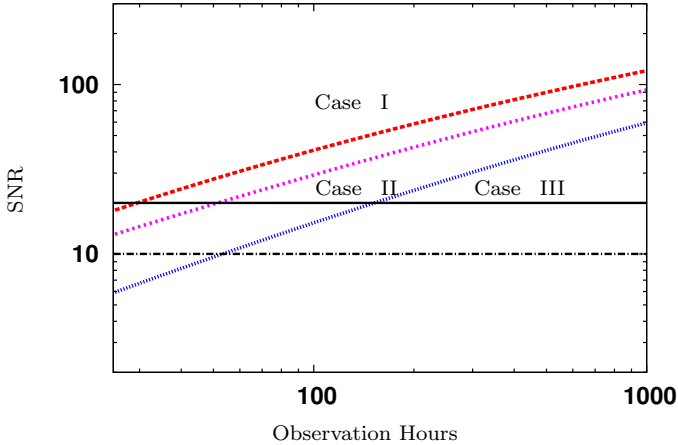}
\caption{The dashed, dotted and fine-dotted lines show the predictions for
  Case I, II and III respectively when all the available $k$-modes are
  combined. The horizontal dot-dashed and solid lines mark the
  SNR of $10$ and $20$ respectively.}
\label{fig:e}
\end{figure}

For comparison, we consider the upcoming SKA-Low which will operate 
within a frequency range of $50 - 350 \, {\rm MHz}$ and investigate the prospects 
of detecting the redshifted \HI 21-cm signal power spectrum from Cosmic Dawn.
SKA-Low is expected to be an array of $\sim 513$ 
stations, each of diameter $D = 35 \, {\rm m}$ . Modelling each station
as having a circular aperture of diameter $35 \, {\rm m}$ we estimate
the primary beam pattern to have a full-width at half-maxima (FWHM) of
$6.44^{\circ}$, and we model the primary beam pattern as a Gaussian
$A(\theta) = e^{-(\theta/\theta_0)^2}$ with $\theta_0 = 0.62 \,
\theta_{\rm FWHM} = 4^{\circ}$ \citep{Choudhuri2014}. Using this we have
calculated $\tilde{\eta} = 39.44$ at the frequency of our interest.
We have simulated the SKA-Low baseline distribution corresponding to
$8 \, {\rm hrs}$ of observation towards a fixed pointing direction at
declination $\delta = -30^{\circ}$. We have used the proposed SKA-Low\footnote{\href {https://astronomers.skatelescope.org/wp-content/uploads/2016/09/SKA-TEL-SKO-0000422_02_SKA1_LowConfigurationCoordinates-1.pdf}{SKA1-
LowConfigurationCoordinates-1.pdf}} antenna
configuration for our simulation. The rest of the
analysis was carried out along exactly the same lines as that for the
uGMRT. Note that the grid spacing $\Delta \ppk$ was scaled appropriately
to account for the different value of the diameter $D$, whereas the
value of $\Delta \kpar$ was maintained the same as for the uGMRT.
Figure~\ref{fig:d} shows a comparison between the dimensionless 
\HI 21-cm signal power spectrum and  the corresponding  $1-\sigma$ error. 
For Case I we find that $\Delta^2_{\rm \HI}$ can be
detected with $100$ hours of observation for Fourier modes 
$0.02 < k < 1.0 \, {\rm Mpc}^{-1}$. However, a detection of $10\Delta^2_{\rm 
\HI}$ and $100\Delta^2_{\rm \HI}$ will require less observation time. In Case
II, it is possible to detect $\Delta^2_{\rm \HI}$ in the $k$-range,
$0.1 < k < 1.0 \, {\rm 
Mpc}^{-1}$ and for the pessimistic scenario, {\it i.e.} Case III, we find that 
$\Delta^2_{\rm \HI}$ can be detected in $100$ hours of observation in the 
$k$-range $0.2 < k < 1.0 \, {\rm Mpc}^{-1}$. This results are significantly
promising when compared to the uGMRT. It is a direct consequence of the
fact that SKA-Low has larger number of antennas and better $uv$-coverage
when compared to uGMRT. It is interesting to note that for SKA-Low the
$1-\sigma$ errors
do not decrease very much as the observing time is increased from $100$ to
$1000 \, {\rm hrs}$,   whereas we find a pretty significant decrease for
uGMRT (Figure~\ref{fig:b}). This indicates that the $1-\sigma$ errors for
the SKA-Low are largely cosmic variance dominated whereas these are the
system noise dominated for the uGMRT.
Figure~\ref{fig:e} shows the predicted SNR as a function of the observing time for SKA-Low.
 The horizontal solid and the dot-dashed lines mark the SNR value of $20$
and $10$ respectively. For Case I, II and III we find that a $20\sigma$ detection
$\Delta^2_{\rm  \HI}$ is possible in $\sim 40,\, 60 \, {\rm and} \, 200$ hours of
observation respectively. Here again it is interesting to note that for
SKA-Low the SNR does not increase as rapidly with the observing time
as for the uGMRT (Figure~\ref{fig:c}). As noted earlier, this is a
consequence of the fact that for SKA-Low the cosmic variance makes a larger
contribution to the total error budget as compared to the uGMRT. Our
predictions for SKA-Low are roughly consistent with the earlier predictions
of \citet{Koopmans2015}.

The upcoming experiment, the Hydrogen Epoch of 
  Reionization Array (HERA; \citealt{DeBoer2017}) a 350-element
  interferometer  is also expected to operate from $50$ to $250
  \, {\rm MHz}$. Note that HERA is a drift scan instrument unlike
  uGMRT and SKA-Low which can track a fixed pointing direction
  on the sky. Considering $1000 \, {\rm hrs} $ of observation, 
at $z = 17$ the HERA sensitivity is about $8$ times better for 
the cosmic dawn \HI 21-cm power spectrum measurement
\citep{DeBoer2017} as compared to uGMRT.

\section{Conclusion} If the proposed b-DM interaction
\citep{Barkana2018,Fialkov2018} enhances the Cosmic Dawn \HI 21-cm
power spectrum, it can be detected with the Band-1 of uGMRT within
reasonable hours of observation. Such a detection would be an
independent confirmation of the enhanced dip reported by
\citet{Bowman2018}. 
The b-DM scattering model depends upon 
two additional parameters: the mass of the DM particles, $0.0032 < m_{\chi} 
< 100 \, {\rm GeV}$ and the cross-section $10^{-30} < \sigma_1 < 3.16 \times 
10^{-18} \, {\rm cm}^2$. This interaction model enhances the Cosmic Dawn \HI 21-cm 
brightness temperature fluctuations and the maximal fluctuation amplitude 
(due to b-DM scattering only) is predicted to be between $0$ and $850 
\, {\rm mK}$, while the maximal fluctuation amplitude without b-DM scattering 
has been predicted to be anywhere between $1.5$ and $90 \, {\rm mK}$ \citep{Fialkov2018}.
With the measurement of the Cosmic Dawn \HI 21-cm power spectrum,
one expects to constrain the $m_{\chi} - \sigma_1$ parameter space.
Observations with the Band-1 of uGMRT hold the
prospect of being an interesting probe of the b-DM interaction in the
early universe. Even upper limits from a non-detection of this power
spectrum would impose useful constraints on the mass of the DM particles, 
the scattering cross-section and the proposed b-DM interaction. We also note that the 
observations with the Band-1 of uGMRT hold the possibility to constraint 
the ``minicharged” dark matter models.
Even an upper limit from a non-detection of the 
the Cosmic Dawn \HI 21-cm power spectrum can put an upper limit on such 
``minicharged" dark matter models. For example, considering $1000 \, {\rm hrs}$ of observation
with Case I, uGMRT would be sensitive enough to rule out models with
the fraction of ``minicharged" dark matter particles in the range
$f_{dm} \geq 0.03$ based on the \HI 21-cm power spectrum predictions of 
\citet{Munoz2018b}.

{\it Acknowledgement:} The authors would to thank Ravi Subrahmanyan
for drawing their attention to the possibility of observing the Cosmic
Dawn redshifted 21-cm signal using Band-1 of uGMRT.  The authors would
also like to thank Abinash K. Shaw and Anjan K. Sarkar for useful
discussions. SC acknowledges the University Grants Commission, India
for providing financial support through Senior Research Fellowship.

\bibliographystyle{mnras} \bibliography{mylist.bib}

\begin{thebibliography}{34}
\expandafter\ifx\csname natexlab\endcsname\relax\def\natexlab#1{#1}\fi

\bibitem[{Ali} et~al.(2008){Ali}, {Bharadwaj} \& {Chengalur}]{Ali2008}
{Ali} S.~S., {Bharadwaj} S., {Chengalur} J.~N., 2008, \mnras, 385, 2166

\bibitem[Barkana(2018)]{Barkana2018}
Barkana R., 2018, Nature, 555, 7694, 71

\bibitem[Bernardi et~al.(2016)Bernardi, Zwart, Price et~al.]{Bernardi2016}
Bernardi G., Zwart J. T.~L., Price D., et~al., 2016, Monthly Notices of the
  Royal Astronomical Society, 461, 3, 2847

\bibitem[{Bharadwaj} \& {Ali}(2005)]{Bharadwaj2005}
{Bharadwaj} S., {Ali} S.~S., 2005, \mnras, 356, 1519

\bibitem[Bowman et~al.(2018)Bowman, Rogers, Monsalve, Mozdzen \&
  Mahesh]{Bowman2018}
Bowman J.~D., Rogers A. E.~E., Monsalve R.~A., Mozdzen T.~J., Mahesh N., 2018,
  Nature, 555, 7694, 67

\bibitem[Chengalur et~al.(2007)Chengalur, Gupta \& Dwarkanath]{Chengalur2007}
Chengalur J.~N., Gupta Y., Dwarkanath K., 2007, Low frequency radio astronomy
  3rd edition

\bibitem[{Choudhuri} et~al.(2014){Choudhuri}, {Bharadwaj}, {Ghosh} \&
  {Ali}]{Choudhuri2014}
{Choudhuri} S., {Bharadwaj} S., {Ghosh} A., {Ali} S.~S., 2014, \mnras, 445,
  4351

\bibitem[Cohen et~al.(2017)Cohen, Fialkov, Barkana \& Lotem]{Cohen2017}
Cohen A., Fialkov A., Barkana R., Lotem M., 2017, Monthly Notices of the Royal
  Astronomical Society, 472, 2, 1915

\bibitem[{Datta} et~al.(2010){Datta}, {Bowman} \& {Carilli}]{Datta2010}
{Datta} A., {Bowman} J.~D., {Carilli} C.~L., 2010, \apj, 724, 526

\bibitem[DeBoer et~al.(2017)DeBoer, Parsons, Aguirre et~al.]{DeBoer2017}
DeBoer D.~R., Parsons A.~R., Aguirre J.~E., et~al., 2017, Publications of the
  Astronomical Society of the Pacific, 129, 974, 045001

\bibitem[Dvorkin et~al.(2014)Dvorkin, Blum \& Kamionkowski]{Dvorkin2014}
Dvorkin C., Blum K., Kamionkowski M., 2014, Phys. Rev. D, 89, 023519

\bibitem[Ewall-Wice et~al.(2018)Ewall-Wice, Chang, Lazio, Dore, Seiffert \&
  Monsalve]{Ewall-Wice2018}
Ewall-Wice A., Chang T.~C., Lazio J., Dore O., Seiffert M., Monsalve R.~A.,
  2018, arXiv:1803.01815

\bibitem[Feng \& Holder(2018)]{Feng2018}
Feng C., Holder G., 2018, The Astrophysical Journal Letters, 858, 2, L17

\bibitem[Fialkov et~al.(2018)Fialkov, Barkana \& Cohen]{Fialkov2018}
Fialkov A., Barkana R., Cohen A., 2018, arXiv:1802.10577

\bibitem[Gupta et~al.(2017)Gupta, Ajithkumar, Kale et~al.]{Gupta2017}
Gupta Y., Ajithkumar B., Kale H., et~al., 2017, CURRENT SCIENCE, 113, 4, 707

\bibitem[Koopmans et~al.(2015)]{Koopmans2015}
Koopmans L. V.~E., et~al., 2015, PoS, AASKA14, 001

\bibitem[Mellema et~al.(2013)Mellema, Koopmans, Abdalla et~al.]{Mellema2013}
Mellema G., Koopmans L. V.~E., Abdalla F.~A., et~al., 2013, Experimental
  Astronomy, 36, 1, 235

\bibitem[{Morales} et~al.(2012){Morales}, {Hazelton}, {Sullivan} \&
  {Beardsley}]{Morales2012}
{Morales} M.~F., {Hazelton} B., {Sullivan} I., {Beardsley} A., 2012, \apj, 752,
  137

\bibitem[Mu\~noz et~al.(2018)Mu\~noz, Dvorkin \& Loeb]{Munoz2018b}
Mu\~noz J.~B., Dvorkin C., Loeb A., 2018, arXiv:1804.01092

\bibitem[Mu\~noz et~al.(2015)Mu\~noz, Kovetz \& Ali-Ha\"{\i}moud]{Munoz2015}
Mu\~noz J.~B., Kovetz E.~D., Ali-Ha\"{\i}moud Y., 2015, Phys. Rev. D, 92,
  083528

\bibitem[Mu\~noz \& Loeb(2018)]{Munoz2018a}
Mu\~noz J.~B., Loeb A., 2018, Nature, 557, 7707, 684

\bibitem[{Parsons} et~al.(2012){Parsons}, {Pober}, {Aguirre}, {Carilli},
  {Jacobs} \& {Moore}]{Parsons2012b}
{Parsons} A.~R., {Pober} J.~C., {Aguirre} J.~E., {Carilli} C.~L., {Jacobs}
  D.~C., {Moore} D.~F., 2012, \apj, 756, 165

\bibitem[{Philip} et~al.(2018){Philip}, {Abdurashidova}, {Chiang}
  et~al.]{Philip2018}
{Philip} L., {Abdurashidova} Z., {Chiang} H.~C., et~al., 2018,
  {arXiv:1806.09531}

\bibitem[{Pober} et~al.(2014){Pober}, {Liu}, {Dillon} et~al.]{Pober2014}
{Pober} J.~C., {Liu} A., {Dillon} J.~S., et~al., 2014, \apj, 782, 66

\bibitem[{Pritchard} \& {Loeb}(2012)]{Pritchard2012}
{Pritchard} J.~R., {Loeb} A., 2012, Reports on Progress in Physics, 75, 8,
  086901

\bibitem[Santos et~al.(2010)Santos, Ferramacho, Silva, Amblard \&
  Cooray]{Santos2010}
Santos M.~G., Ferramacho L., Silva M.~B., Amblard A., Cooray A., 2010, Monthly
  Notices of the Royal Astronomical Society, 406, 4, 2421

\bibitem[Shankar et~al.(2009)Shankar, Dwarakanath, Amiri et~al.]{Shankar2009}
Shankar N.~U., Dwarakanath K., Amiri S., et~al., 2009, The Low-Frequency Radio
  Universe, 407, 393

\bibitem[Singh et~al.(2017)Singh, Subrahmanyan, Shankar et~al.]{Singh2017}
Singh S., Subrahmanyan R., Shankar N.~U., et~al., 2017, The Astrophysical
  Journal Letters, 845, 2, L12

\bibitem[{Swarup} et~al.(1991){Swarup}, {Ananthakrishnan}, {Kapahi}, {Rao},
  {Subrahmanya} \& {Kulkarni}]{Swarup1991}
{Swarup} G., {Ananthakrishnan} S., {Kapahi} V.~K., {Rao} A.~P., {Subrahmanya}
  C.~R., {Kulkarni} V.~K., 1991, Current Science, Vol.~60, NO.2/JAN25, P.~95,
  1991, 60, 95

\bibitem[Tashiro et~al.(2014)Tashiro, Kadota \& Silk]{Tashiro2014}
Tashiro H., Kadota K., Silk J., 2014, Phys. Rev. D, 90, 083522

\bibitem[Trott et~al.(2012)Trott, Wayth \& Tingay]{Trott2012}
Trott C.~M., Wayth R.~B., Tingay S.~J., 2012, The Astrophysical Journal, 757,
  1, 101

\bibitem[{Vedantham} et~al.(2012){Vedantham}, {Udaya Shankar} \&
  {Subrahmanyan}]{Vedantham2012}
{Vedantham} H., {Udaya Shankar} N., {Subrahmanyan} R., 2012, \apj, 745, 176

\bibitem[Voytek et~al.(2014)Voytek, Natarajan, García, Peterson \&
  López-Cruz]{Voytek2014}
Voytek T.~C., Natarajan A., García J. M.~J., Peterson J.~B., López-Cruz O.,
  2014, The Astrophysical Journal Letters, 782, 1, L9

\bibitem[Xu et~al.(2018)Xu, Dvorkin \& Chael]{Xu2018}
Xu W.~L., Dvorkin C., Chael A., 2018, Phys. Rev. D, 97, 103530

\end{thebibliography}
\end{document}